\documentclass[preprint]{aastex}

\newcommand{\be}	{\begin{equation}}
\newcommand{\ee}	{\end{equation}}
\newcommand{\beq}	{\begin{eqnarray}}
\newcommand{\eeq}	{\end{eqnarray}}
\newcommand{\bd}	{\begin{displaymath}}
\newcommand{\ed}	{\end{displaymath}}
\newcommand{\define}	{\equiv}

\newcommand{\LCDM}	{$\Lambda$CDM~}
\newcommand{\eV}	{\mbox{eV}}
\newcommand{\GeV}	{\mbox{GeV}}
\newcommand{\br}	{{\bf r}}
\newcommand{\bp}	{{\bf p}}
\newcommand{\bq}	{{\bf q}}
\newcommand{\vol}	{{\cal V}}

\begin{document}

\title{\bf Repulsive Dark Matter}

\author{Jeremy Goodman}
\affil{Princeton University Observatory, Princeton NJ 08544}
\email{jeremy@astro.princeton.edu}

\begin{abstract}
It seems necessary to suppress, at least partially,
the formation of structure on subgalactic scales.
As an alternative to warm or collisional dark matter, I postulate
a condensate of massive bosons interacting via a repulsive interparticle
potential, plus gravity.  This leads to
a minimum lengthscale for bound objects, and to superfluidity.
Galactic dynamics may differ significantly
from that of more generic dark matter in not unwelcome ways, especially
in the core.
Such particles can be realized as quanta of a relativistic massive
scalar field with a quartic self-interaction.  At high densities,
the equation of state has the same form as that of an ideal relativistic
gas despite the interactions.  If the nonrelativistic lengthscale
is of order a kiloparsec, then the energy density in these particles
was comparable to that of photons at early times, but small enough
to avoid conflict with primordial nucleosynthesis.

\end{abstract}

\maketitle



\section{Introduction}\label{INTRODUCTION}

At the time of writing,
the combination (``\LCDM'') of cold dark matter, a Harrison-Zeldovich
spectrum of initial fluctuations, and a small
cosmological constant appears to be in good agreement
with many observational constraints: the angular power spectrum of
the temperature of the cosmic microwave background,
the large-scale distribution of galaxies and clusters, and the growth
of present-day structures by gravitational instability 
\citep[and references therein]{BOPS}.
Recently, however, it has become clear that \LCDM overpredicts
structure on small scales.  N-body and hydrodynamic simulations of
galaxy formation within the \LCDM model predict that the dark halos
of galaxies should have singular cores \citep{NFW,NS}, contrary to observation
\citep{Carignan,Swaters}.
A separate though related problem is that late accretion of
dark matter clumps or satellite galaxies
is likely to shred the disks of spirals \citep{TO92,Klypin}.

Clustering on small scales could be suppressed by
an upper limit to the phase-space density of the dark-matter particles
due either to thermal entropy or, if they are fermions, 
degeneracy pressure; the particle mass would have to lie in the range
$10^2~\eV< m<10^3~\eV$ \citep{HD}.

Another proposal is that the dark-matter
particles are heavy ($m\gg\GeV$) but self-interacting \citep{SS}.
Since the simulations tend to produce a dark-matter velocity
dispersion that decreases towards the center of the cusp,
occasional collisions ``heat'' the central regions and produce
a finite-density core, at least temporarily.  Once the inner
regions become roughly isothermal, the core will recollapse;
the collision cross section must be chosen so that this does not
happen within the age of the galaxy.  This proposal seems unlikely
to alleviate the second problem, \emph{viz.}
the damage to galactic disks.

This paper proposes a third solution to the problem of
small-scale power: a Bose-Einstein condensate of dark matter
particles, hence similar to the axion, but interacting \emph{via} a
repulsive potential of finite range.  I show in \S2 that cores would
have a minimum size independent of their density or mass, and that the
dark matter would behave as a superfluid.  Such particles arise fairly
naturally as quanta of a self-interacting scalar field (\S3).  For
plausible choices of the minimum core size, the interaction energy per
particle would have been comparable to the rest mass somewhat before
the universe became matter-dominated, and at earlier times it would
have a relativistic equation of state ($p\approx \rho/3$), hence
slightly increasing the effective number of degrees of freedom in the
radiation field, but not enough to violate the constraints of
primordial nucleosynthesis.  With respect to the standard cold-dark
matter spectrum, density fluctuations in the linear regime would be
suppressed on comoving scales less than a few megaparsecs (\S4).

While this work was being written up, I became aware that my
colleague P. J. E. Peebles has been working along similar lines
\citep{PV,P99,P00}.  There is also closely related earlier work
\citep{Tkachev85,Tkachev91}.  This
seems to be a genuine case of convergent evolution, and the fact that
independent groups arrived at similar results is perhaps reassuring in
such a wildly speculative domain.
I started with a nonrelativistic
many-body view of these particles (\S\ref{NONREL}) and then sought a
relativistic framework for them (\S{\ref{REL}), whereas \citet{PV}
seem to have proceeded in the opposite direction.  With further
development, the nonrelativistic but fully quantum-mechanical viewpoint
may prove useful in studying the galactic dynamics of this form of dark 
matter.

\goodbreak
\section{Minimum dark-matter core radius}\label{NONREL}

Suppose that nonrelativistic bosons interact \emph{via} a
a two-particle potential $U(\br_1-\br_2)$ of finite range.
The potential energy of $N$ such bosons in a common single-particle momentum
state $\psi(\br)=\vol^{-1/2}e^{i\bp\cdot\br}$ in volume $\vol$ is 
\be\label{WN}
W_N = \frac{N(N-1)}{2 \vol}\,\tilde U(0) ~\define~
\frac{N(N-1)}{2 \vol}\,\int d\br'\, U(\br'),
\ee
where $\tilde U(\bp')$ is the fourier transform of $U(\br')$, whose
range is assumed to be small compared to the linear dimensions of $\vol$.
If $N,\vol\to\infty$ at fixed number density $n=N/\vol$, then the 
potential energy per unit volume is $w(n)= n^2\tilde U(0)/2$.
For the moment,  $w(n)/n\ll mc^2$, the boson rest-mass energy.

Macroscopically, one has a polytropic gas of adiabatic
index $\gamma=2$; in other words, the pressure is related to
the mass density $\rho\approx mn$ by
\be\label{EOS_NR}
p= K\rho^2,\qquad K= \tilde U(0)/2m^2.
\ee
If the gas is self-gravitating, there is a well-known spherical
equilibrium with the density profile \citep{Chandra}:
\be\label{EMDEN}
\rho(r)=\rho(0)\frac{\sin(r/a)}{r/a},\qquad a= \sqrt{\frac{K}{2\pi G}}.
\ee
The radius of the sphere, $\pi a$, is independent of
the central density $\rho(0)$, which determines the total mass,
$4\pi^2 a^3\rho(0)$.
Dark-matter halos do not have the profile (\ref{EMDEN}), but
this does not rule out the model.
If the particles are not all in the same momentum state,
then their relative motions make an additional contribution to the pressure,
which allows the halo to have a power-law density profile outside the
core.  Indeed, axionic dark matter is usually assumed to be a Bose-Einstein
condensate like the one considered here but without the repulsive interaction,
so the pressure support of dark halos in that model is due entirely to
relative motions.
Still, since (\ref{EOS_NR}) gives the minimum pressure, (\ref{EMDEN})
is the most compact possible equilibrium.
The conventional definition of the the core radius in terms of
the central density and its second derivative is
$r_c\define \left[-3\rho(0)/\rho''(0)\right]^{1/2}$, thus 
$r_{c,\min}=3a$.

The nonrelativistic approximation breaks down when the pressure
(\ref{EOS_NR}) is comparable to the rest mass density.  This happens
when $\rho\approx c^2/Ga^2\define \rho_{\rm rel}$.  If this bosonic
dark matter dominates the total present-day mass density,
then the mean mass density was equal to
$\rho_{\rm rel}$ at redshift
\be\label{Z_REL}
1+z_{\rm rel} = \left(\frac{8\pi G\rho_{\rm rel}}{3\Omega H_0^2}\right)^{1/3}
\approx 2.1\times 10^5~\Omega_{0.3}^{-1/3}
h_{50}^{-2/3} r_{\rm c,kpc}^{-2/3},
\ee
where $\Omega\define\Omega/0.3$,
$h_{50}\equiv H_0/(50~{\rm km~s^{-1}~Mpc^{-1}})$, and
$r_{\rm c,kpc}=r_{\rm c,min}/{\rm kpc}$.
The question how the dark matter behaves at higher redshifts
is deferred to \S\S3-4.  It suffices for now that
$1+z_{\rm rel}$ is comfortably larger than the redshift of
matter-radiation equality \citep{PPC}:
$1+z_{\rm eq}= 1.8\times 10^3\,\Omega_{0.3} h_{50}^2$.

The interaction makes the gas a superfluid.
If one particle is removed from the Bose-Einstein condensate
and put into a single-particle state with momentum $\bp+\bq\ne\bp$,
then the potential energy (\ref{WN}) is replaced by
\bd
W_{N-1} + \frac{N-1}{\vol}\left[\tilde U(0)+\tilde U(\bq)\right]
~=W_{N} + \frac{N-1}{\vol}\,\tilde U(\bq).
\ed
The first term on the left
is the interaction of the $N-1$ particles
in the condensate with one another, and the second is the interaction
of the condensate with the extracted particle; the piece
involving $\tilde U(\bq)$ is the exchange energy resulting from
symmetrization of the $N$-particle wavefunction.
I assume that the range of $U(\br)$ is sufficiently short so that
$\tilde U(\bq)\approx\tilde U(0)$.
The energetic penalty for removing a particle from the condensate
is then approximately equal to the potential energy per particle pair,
$n\tilde U(0)$.  Thus if the condensate streams past an
obstacle (an external potential) at speed $v$,
scattering out of the condensate is impossible if the kinetic
energy per particle  is less than this energy penalty, \emph{i.e.}
if the relative velocity
\be\label{MAXP}
v < \sqrt{2 n \tilde U(0)/m}\equiv v_{\rm crit}(n).
\ee
Similarly, when two condensates of density $n_{1,2}$ and
momenta $\bp_{1,2}$ stream through one another, dissipation occurs
only if $|\bp_1-\bp_2|\ge \sqrt{2}m v_{\rm crit}(n_1+n_2)$.  This is not to
say that the two streams do not interact, but rather that they interact
only through the mean-field energy $n_1n_2\tilde U(0)$ per unit volume.
Inside a core supported mainly by the repulsive interaction 
[eq.~(\ref{EMDEN})], $v_{\rm crit}$ is comparable to the virial velocity.

It will be interesting to study whether superfluid dark matter would
have any distinctive consequences for galactic dynamics other than the
minimum core size.  Attention naturally focuses on dissipative
processes, such as dynamical friction: \emph{i.e.}, irreversible
transfer of energy and momentum between the dark and baryonic matter
\emph{via} by their gravitational interaction \citep[cf.][]{BT}. In
collisionless systems, dynamical friction involves upon
single-particle resonances \citep[e.g.][]{TW}, much like Landau
damping in plasmas.  As long as the condition (\ref{MAXP}) is
satisfied, however, a perturbing gravitational potential interacts
coherently with the condensate, and all particles have the same
resonant frequencies because they share a common macroscopic
wavefunction.  Thus for example, a rotating galactic bar may
experience little drag against the dark matter; this may circumvent an
important argument against dense dark halos in barred spirals
\citep{Sellwood}.
The question will require a quantitative analysis, however, because
even in the innermost parts parts of the galaxy, not all of the
dark matter will be in the condensate.

\goodbreak
\section{Relativistic era}\label{REL}

Dark matter with the properties described in \S\ref{NONREL}
arise as quanta of a self-interacting relativistic
scalar field $\phi$ with lagrangian density
\be\label{LAGR}
{\cal L}=-\sqrt{-g}\left(\frac{1}{2}g^{\mu\nu}\partial_\mu\phi\partial_\nu\phi
~+V(\phi)\right).
\ee
Without loss of generality, the minimum of $V(\phi)$ occurs at $\phi=0$,
and $V(\phi)=m^2\phi^2/2 + \mbox{(higher powers)}$.
Potentials of the form
\be\label{VPHI}
V(\phi) = \frac{1}{2}m^2\phi^2 +\kappa\phi^4
\ee
are of particular interest, though
one might want to add a constant $V(0)=\Lambda/8\pi G$
to produce a present-day cosmological constant.
In lowest-order perturbation theory,
the interaction energy of a state $|\Psi_N(0)\rangle$ consisting of
$N$ quanta at rest in volume $\vol$ is, in Minkowski space,
\be\label{PERT}
\int\limits_{\vol} d^3\br\, \langle\Psi_N(0)|\,{\bf :}\,
\kappa \phi^4(\br,t)\,{\bf :}\,
|\Psi_N(0)\rangle = \frac{6\kappa}{(2m)^2}\,\frac{N(N-1)}{\vol},
\ee
and therefore $\tilde U(0) = 3\kappa/m^2$.

Semiclassical methods give the same result, which is 
important because they are not restricted
to perturbation theory.  Thus if $\phi$ were a spatially uniform
classical field, then ${\cal L}$ could be regarded as the lagrangian
of a one-dimensional oscillator with an explicit time dependence
\emph{via} the metric $g_{\mu\nu}\to\mbox{diag}(-1,a^2,a^2,a^2)$
in an Einstein-de Sitter universe.
The momentum conjugate to $\phi$ is $\varpi= a^3\dot\phi$, the
hamiltonian is
\bd
{\cal H}= \frac{\varpi^2}{2a^3} + a^3 V(\phi),
\ed
and the action in the oscillator is given by an integral
over one complete cycle:
\be\label{ACTION}
{\cal I} = \frac{1}{2\pi}\oint \varpi d\phi
~=~\frac{a^3}{\pi\sqrt{2}}\oint \sqrt{a^{-3}{\cal H}- V(\phi)}~d\phi.
\ee
Semiclassically, ${\cal I}$ becomes the number of quanta
per comoving volume, $na^3$,
while ${\cal H}$ becomes the energy per comoving volume, $\rho a^3$.
(In this section, $\rho$ will be the total energy density, 
not the rest-mass density $mn$ of \S\ref{NONREL}.)
The definition (\ref{ACTION}) makes sense only when the
oscillation frequency $\omega=(\partial{\cal H}/\partial{\cal I})_a$
is much larger than the current Hubble expansion rate $\dot a/a$, in
which case ${\cal I}$ is an adiabatic invariant and hence $na^3$ is
conserved.  By direct expansion of the quadrature (\ref{ACTION}) to first
order in $\kappa$ and inversion of series, one has
\beq\label{SMALLPHI}
{\cal H} &=& m{\cal I} + \frac{3\kappa}{2m^2 a^3}\,{\cal I}^2
~+ O(\kappa^2{\cal I}^3)\nonumber\\[2ex]
\mbox{hence}\quad \rho &=& mn + \frac{3\kappa}{2m^2}\,n^2
~+ O(\kappa^2 n^3),
\eeq
in agreement with previous results for the nonrelativistic (small-$n$)
regime.
In the opposite limit of large $n$, the quadrature~(\ref{ACTION})
is dominated by  $\phi\gg m/\sqrt{\kappa}$;
neglecting the mass term in $V(\phi)$, one has
\be\label{EOS_REL}
\rho \approx 3^{4/3}\pi^2\Gamma^{-8/3}(1/4)\,\kappa^{1/3} n^{4/3}
\approx 1.377\, \kappa^{1/3}\, n^{4/3},
\ee
as if this were a \emph{noninteracting}
relativistic gas: $p=-\partial(\rho\vol)/\partial\vol = \rho/3$.
Eq.~(\ref{ACTION}) can be evaluated to an exact expression for
$n(\rho)$ in terms of complete elliptic integrals.  

We are now in a position to estimate the mass $m$ and average 
number density $\bar n(z)$ of these quanta.
From eqs.~(\ref{EMDEN}) \& (\ref{SMALLPHI}), it follows that
the minimum core radius $r_{\rm c,min}=3a$ depends only upon
$m^4/\kappa$ and fundamental constants, so
\be
mc^2 = \left(\frac{27\hbar^3 c^3\kappa}{4\pi G r_{\rm c,\min}^2}\right)^{1/4}
\approx 10.7~\kappa^{1/4}\,r_{\rm c,kpc}^{-1/2}
~\eV.
\ee
Apart from the dimensionless coupling $\kappa$, this is the geometric
mean of the Planck mass and the mass whose Compton wavelength is
$2\pi r_{\rm c,\min}$.  Furthermore, if this form of dark matter dominates
the mass density today, then
\be
\bar n(z) = \frac{\Omega\rho_{\rm crit}}{m}(1+z) \approx
74.~\kappa^{-1/4} r_{\rm c,kpc}^{1/2}\Omega_{0.3} h_{50}^2~
(1+z)^3 ~\mbox{cm}^{-3}.
\ee
Prior to the redshift (\ref{Z_REL}) when particles followed
the relativistic equation of state (\ref{EOS_REL}), they would have
contributed a constant fraction of the
total energy density,
equivalent to an increase 
\be\label{DELTANU}
\Delta N_\nu\approx 
0.14 ~r_{\rm c,kpc}^{2/3}\, (\Omega_{0.3} h_{50}^2)^{4/3}
\ee
in the number of effectively massless neutrinos (assuming $N_\nu\approx 3$),
which is compatible
with the constraints from primordial nucleosynthesis \citep{OT}.

\section{DISCUSSION}\label{DISCUSSION}	

We have seen that small-scale structure can be suppressed even if
the dark matter is completely cold and bosonic, provided that it has
a repulsive interaction.  At first blush, the idea seems less
natural than the alternatives---warm or degenerate fermionic
dark matter---which have been much more widely discussed.
I are not aware of a strong particle-physics motivation for matter
with these properties.

Nevertheless, in working through the consequences of the basic
idea, one is intrigued by some satisfying coincidences.
\begin{itemize}
\item[(i)] From a nonrelativistic viewpoint (\S\ref{NONREL}),
the equation of state (\ref{EOS_NR}) results from a generic
two-body interaction of finite range among massive bosons;
relativistically, it emerges from the simplest nonlinear field
theory (\ref{LAGR})-(\ref{VPHI}).

\item[(ii)] The nonrelativistic equation of state implies a characteristic
lengthscale and a minimum dark-halo core radius.
If this lengthscale is of order a kiloparsec, as the observations
suggest \citep{HD}, then the dark matter began to be nonrelativistic
at the lowest possible redshift that growth of structure would permit,
\emph{viz.} $z_{\rm rel}\sim z_{\rm eq}$. The result (\ref{DELTANU})
that the energy density in these hypothetical quanta would have
been comparable to the energy density in photons at early
times is really the same coincidence.  Both are independent of $\kappa$
and $m$, because the relevant combination of these quantities is already
fiexed by $r_{\rm c,min}$ \& $\Omega$.

\end{itemize}

With regard to the second point, \cite{P00} has estimated that
the model may be a little too successful at suppressing small-scale
power during the linear regime.  Density fluctuations that
come within the horizon before $z_{\rm eq}$ not only do not grow,
but actually decay, until their physical size is larger than
the ``Jeans length'' $a$.  He finds that this constraint is marginally
inconsistent with the quartic model unless $r_{\rm c,min}\le 0.5~{\rm kpc}$.
Pending more precise observations, however, one may be impressed that the
model marginally survives this test without appeal to an adjustable
parameter.\footnote{Peebles notes that slightly sub-quartic potentials
$V(\phi)=m^2\phi^2/2 + \kappa|\phi|^q$ with $q\approx 3.7$
would fit this constraint more comfortably.  But one would have
to sacrifice (i).}

\acknowledgments
I thank David Spergel for provoking my interest in dark matter,
Paul Steinhardt for technical advice, and especially Jim Peebles for 
wide-ranging discussions.


\end{document}